\documentclass{article}
\usepackage{mathptmx}

\usepackage{float}
\usepackage{amssymb, amsmath, amsthm}
\usepackage{mathrsfs}
\usepackage{tabulary}
\usepackage{tabularx}
\usepackage{rotating}

\usepackage{setspace}
\usepackage{multirow}
\usepackage{paralist}
\usepackage{listings}
\usepackage[T1]{fontenc}
\usepackage[utf8]{inputenc}
\usepackage[english]{babel}
\usepackage{dsfont}
\usepackage{hyperref}
\usepackage{subcaption}
\usepackage{amsmath}
\usepackage{graphicx}
\usepackage{url}
\usepackage{booktabs}
\usepackage{pifont}
\usepackage{comment}
\newcommand{\cmark}{\ding{51}}%
\newcommand{\xmark}{\ding{55}}%

\usepackage{authblk}
\usepackage{etoolbox}
\usepackage{longtable}
\usepackage[super,numbers,compress]{natbib}
\usepackage{color}

\usepackage{caption}
\usepackage{hyperref}
\usepackage{subcaption}

\usepackage{censor}

\usepackage[hmargin=2.54cm,vmargin=2.54cm]{geometry}

\newcommand{\E}{\mathsf{E}}

\newcommand{\dd}{\mathrm{d}}

\usepackage{pdflscape}
\usepackage{amsmath, booktabs, caption, longtable}
\usepackage{tabularx} 
\usepackage{threeparttable}
\usepackage{tablefootnote}
\usepackage{enumitem}
\usepackage[hmargin=2.54cm,vmargin=2.54cm]{geometry}
\setlist[itemize]{itemsep=0pt, topsep=0pt, partopsep=0pt, parsep=0pt}
\setlist[enumerate]{itemsep=0pt, topsep=0pt, partopsep=0pt, parsep=0pt}

\usepackage{sectsty} 
\sectionfont{\fontsize{14}{16.8}\selectfont}
\subsectionfont{\fontsize{12}{15.6}\selectfont}
\usepackage{chapterbib}

\title{Everything all at once: 
On choosing an estimand for multi-component environmental exposures 

}
\author{Kara E. Rudolph, Shodai Inose, Nicholas Williams, Ivan Diaz, Lucia Calderon, Jacqueline M. Torres, Marianthi-Anna Kioumourtzoglou}
\date{}

\begin{document}
\maketitle


\begin{abstract}
  Many research questions---particularly those in environmental health---do not involve binary exposures. In environmental epidemiology, this includes multivariate exposure mixtures with nondiscrete components. Causal inference estimands and estimators to quantify the relationship between an exposure mixture and an outcome are relatively few. 
We propose an approach to quantify a relationship between a shift in the exposure mixture and the outcome---either in a single timepoint or longitudinal setting. The shift in the exposure mixture can be defined flexibly in terms of shifting one or more components, including examining interaction between mixture components, and in terms of shifting the same or different amounts across components. The estimand we discuss has a similar interpretation as a main effect regression coefficient. 
First, we focus on choosing a shift in the exposure mixture supported by observed data. We demonstrate how to assess extrapolation and modify the shift to minimize reliance on extrapolation. 
Second, we propose estimating the relationship between the exposure mixture shift and outcome completely nonparametrically, using machine learning in model-fitting. This is in contrast to other common approaches, which employ parametric modeling for at least some relationships, which we would like to avoid because parametric modeling assumptions in complex, nonrandomized settings are tenuous at best. 
We are motivated by longitudinal data on pesticide exposures among participants in the CHAMACOS Maternal Cognition cohort. We examine the relationship between longitudinal exposure to agricultural pesticides and risk of hypertension. 
We provide step-by-step code to facilitate the easy replication and adaptation of the proposed approach.
\end{abstract}

\section{Introduction}\label{sec:intro}

In epidemiology, we are most familiar with estimating the effects of binary treatments or exposures. However, in environmental epidemiology, the ``exposure" is often a mixture of non-discrete (e.g., continuous) components,
\cite{braun2016can} 
and one may be interested in different mixture-outcome associations: 
 1) the association between an individual mixture component and an outcome,
 2) the association between multiple (or all) mixture components jointly and the outcome, and 
 3)  the extent to which mixture components are associated with the outcome in an interactive (e.g., supra-additive or supra-multiplicative) way.\cite{braun2016can,hamra2018environmental,gibson2019complex,mccoy2023semiparametric,smith2023estimating}
We introduce notation to make our discussion concrete. 
The exposure at timepoint $t$ is denoted: $\mathbf{A}_t=\{A_{1,t}, ..., A_{J,t}\}$ for $J$ components of the exposure mixture. One may also be interested in cumulative exposure to a mixture at 
multiple timepoints, $\bar{ \mathbf{A}} = (\mathbf{A}_0, ... \mathbf{A}_T)$ for exposures at times $t \in \{0,... T\}$. This longitudinal setting, though common, is less frequently considered. 
This is a complex exposure: 1) it is longitudinal; 2) at each timepoint, 
the exposure mixture consists of multiple components, 
 3) some or all of which 
 may be continuous or otherwise multi-valued; and 4) interdependent. 

When considering complex exposure mixtures, 
researchers typically estimate effects of point-in-time individual, joint, summary, or interacting mixture components using: 
i) Bayesian approaches (e.g., Bayesian Kernel Machine Regression\cite{bobb2015bayesian}, Bayesian Multiple Index Models\cite{mcgee2023bayesian}), ii) parametric regression models (e.g., Cox proportional hazards models)\cite{papadogeorgou2019low}---possibly using regularization (e.g., lasso, partial least squares\cite{hamra2018environmental}), or iii) estimating the effects of hypothetically shifting quantiles of a summary exposure index, effectively treating the aggregate index as a single exposure, e.g., weighted quantile sum regression,\cite{hamra2018environmental} quantile g-computation\cite{goin2021hyperlocalized,riddell2021hyper}). There are several comprehensive and recent reviews.\cite{joubert2025workflow,joubert2022powering,hamra2018environmental,braun2016can,gibson2019complex}

Although a growing literature aims to improve estimation of the health 
 effects of complex exposure mixtures  e.g.,~\cite{antonelli2024causal,shin2025treatment,wilson2018model}, several gaps persist in the applied literature that may undermine the robustness of findings.\cite{carone2020pursuit}  
 First, there is infrequent or limited assessment of data support for the quantities estimated (though there are exceptions, e.g.,~\cite{snowden2015framing}), resulting in reliance on model-based extrapolation, which would 
 result in bias to the extent that the model is misspecified in the extrapolation region.
 \cite{antonelli2024causal,
bao2025addressing,king2006dangers,zigler2021invited,keil2021keil} 
 Consider a simple scenario with two exposures: manganese fungicides and neonicotinoids, standardized and shown in Figure \ref{fig:extrapolation} (we discuss the data source below). Shifting a single component of the mixture may seem fine; 
 e.g., in Figure \ref{fig:extrapolation}, shifting neonicotinoid values down by 50\% would not result in any extrapolation for that component marginally, which is bounded [0,1]. However, when considering the \textit{joint exposures},  
Figure \ref{fig:extrapolation} shows that 
 many shifted two-dimensional data-points (shown in red) lie outside the observed range of the two-dimensional data (shown in the gray region)---resulting in extrapolation. This reflects the caution given by Antonelli and Zigler\cite{antonelli2024causal}: ``notions of overlap borrowed from the univariate exposure case do not imply overlap in the mixture." Indeed, the risk and extent of extrapolation increases with increasing dimensionality of the mixture.\cite{antonelli2024causal}
 A second limitation 
 is that most applied analyses rely on parametric modeling for at least some of the confounder-exposure-outcome relationships, which 
 is also expected to result in bias to the extent that the parametric model is misspecified.\cite{rudolph2023all,carone2020pursuit,mccoy2023semiparametric} A third limitation is 
a lack of options for 
estimating cumulative effects of longitudinal exposure mixtures\cite{joubert2024environmental,carone2020pursuit} while appropriately accounting for time-varying exposure-confounder feedback.\cite{robins1986new} 

\begin{figure}[H] 
\captionsetup{justification=raggedright,singlelinecheck=false}
  \caption{Example of extrapolation (2-dimensional red points in the gray shaded region) outside of the convex hull (defined in Section \ref{sec:convexhull} black dashed polygon) when decreasing neonicotinoid values by 50\% while keeping manganese-containing fungicide values as observed. If we considered neonicotinoid as a single pesticide exposure and applied a reductive 50\% shift, we would not extrapolate beyond the bounds of the observed \textit{neonicotinoid} values, as there is support between scaled values of 0 and 1. However, when considering neonicotinoid and manganese-containing fungicide values \textit{jointly} and applying a reductive 50\% shift on the former alone, we find numerous 2-dimensional data points that would extrapolate beyond the bounds of the observed data. In this simple example, we find that there is little support for high-levels of manganese-containing fungicide exposure and simultaneously low-levels of neonicotinoid exposure. Data source: CHAMACOS Maternal Cognition Study}
\centering
\vspace{-1.2cm}
\includegraphics[width=.6\textwidth]{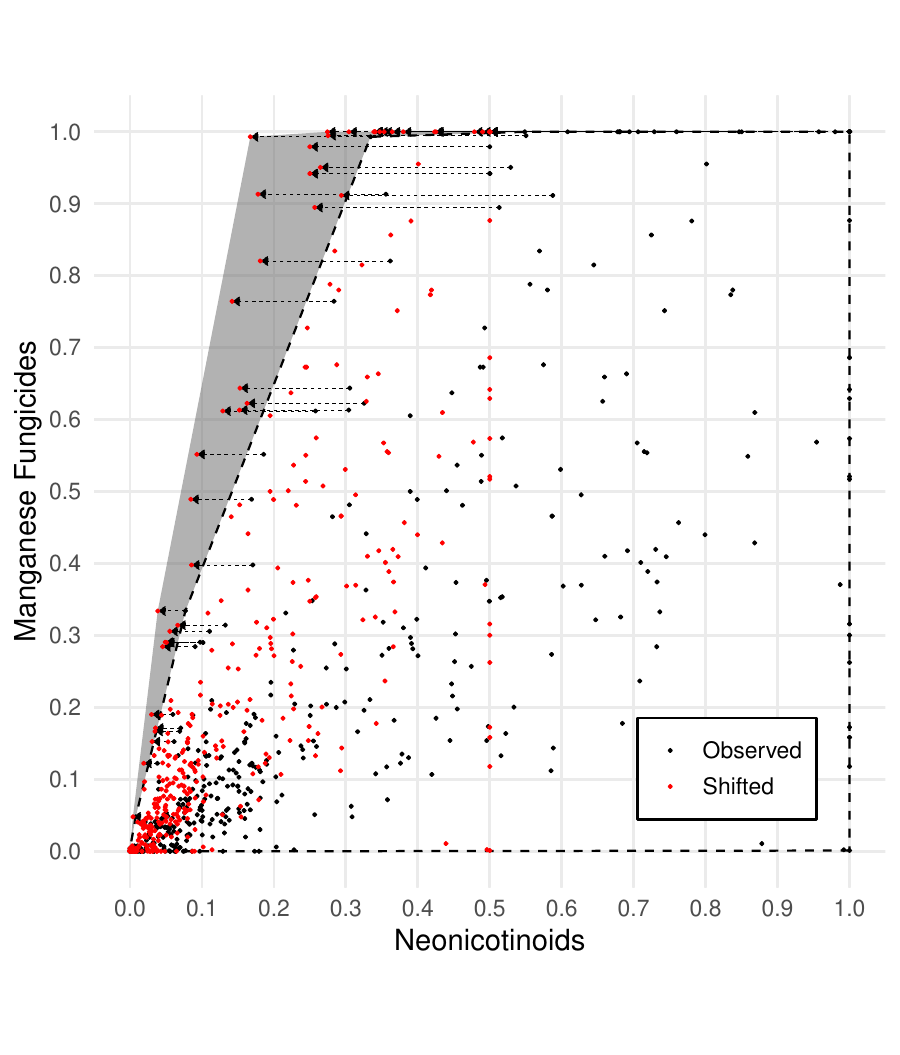}
\label{fig:extrapolation}
\end{figure}

We seek to address these gaps by providing user-friendly, step-by-step code as well as 
open-source software to guide applied researchers through: 1) formulating a causal estimand of the effect of a hypothetical shift intervention on complex, \textit{time-varying environmental mixture} on a \textit{time-varying outcome}; 2) evaluating support for the desired mixture shift (using novel, open-source software) and iterating on the particular hypothetical mixture shift until it is well-supported by the observed data to minimize reliance on extrapolation; and 3) estimating the better-supported quantity using a flexible, nonparametric longitudinal estimator that incorporates machine learning in model fitting to obviate 
reliance on correct parametric model specification and that is computationally scalable to big-data scenarios. 
Our purpose is to 
lower barriers for using existing methods that are rarely used in mixture applications. 
The approach we step through is general and, in principle, can be used for: i) any number of discrete exposure timepoints,
           ii) any number of exposure mixture components, each distributed in any manner, and related to each other in any way, 
           iii) any number of potential confounding variables, time-invariant and time-varying, distributed in any way, and related to other variables in any way,
           iv) time-invariant or time-varying outcomes, distributed in any way,
            v) censoring, loss-to-follow-up, or competing risks over time, and
            vi) big data (e.g., millions of observations).

As a motivating example, we consider the extent to which pesticide exposure affects risk of hypertension among participants in the CHAMACOS (Center for the Health Assessment of Mothers and Children of Salinas) Maternal Cognition Study cohort (described in \ref{sec:supp:cohort})
.\cite{eskenazi2003chamacos} 
Each participant at each time is assigned an exposure level for each of seven classes of pesticides, $\mathbf{A}_t=\{A_{1,t}, A_{2,t}, A_{3,t}, A_{4,t}, A_{5,t}, A_{6,t}, A_{7,t}\}$, corresponding to organophosphates, pyrethroids, carbamates, and neonicotinoids (all insecticides), manganese-containing fungicides, and glyphosate and paraquat (herbicides), 
measured at five timepoints, $\bar{\mathbf{A}} = (\mathbf{A}_0, ..., \mathbf{A}_4)$. 
We refer to pesticide ``exposures" for simplicity (as done previously\cite{keil2021bayesian}); they are the concentration applied 
within 1-kilometer buffers of the participant's address 
(\ref{sec:supp:measures} provides more detail).
We consider relationships between the outcome of hypertension risk by the end of follow-up, $Y$, and: 1) 
exposure to this pesticide mixture at baseline, $\mathbf{A}_0$, and 2) 
longitudinal exposure to this pesticide mixture $\bar{\mathbf{A}}$. 
Our motivating example is just that---an example; we use it to illustrate the challenges and contributions named above; an applied analysis of these data would need to pay additional attention to aspects of the study that are outside the scope of this paper, like the measurement and assignment of the exposure mixture to study participants. 
 


We organize this 
paper around a series of questions an analyst may ask when formulating and estimating quantifiable relationships between 
a 
possibly longitudinal exposure mixture and an outcome. 

\section{Question: What type of quantifiable relationship between the exposure mixture and outcome (i.e., statistical estimand) do we formulate?}

We want to retain full information on the mixture components (i.e., we do not want to discretize continuous components, as this introduces measurement error, which can induce bias---even away from the null\cite{robins1987addendum}---and increase variance.\cite{keogh2020stratos}), and we want to consider all components of the mixture, as they are frequently interdependent.  
Figure \ref{fig:baselinedis} shows the distribution of each pesticide class at baseline ($t=0$) from our motivating example. Figure \ref{fig:baselinecorr} shows correlations between pesticide classes at baseline. 

\begin{figure}[H] 
  \caption{Each pesticide class at baseline, truncated at the 95\% percentile.}
\centering
\begin{subfigure}[t]{0.49\textwidth}
      \caption{Distribution.}
\centering
\vspace{-.6cm}
\includegraphics[width=1\textwidth]{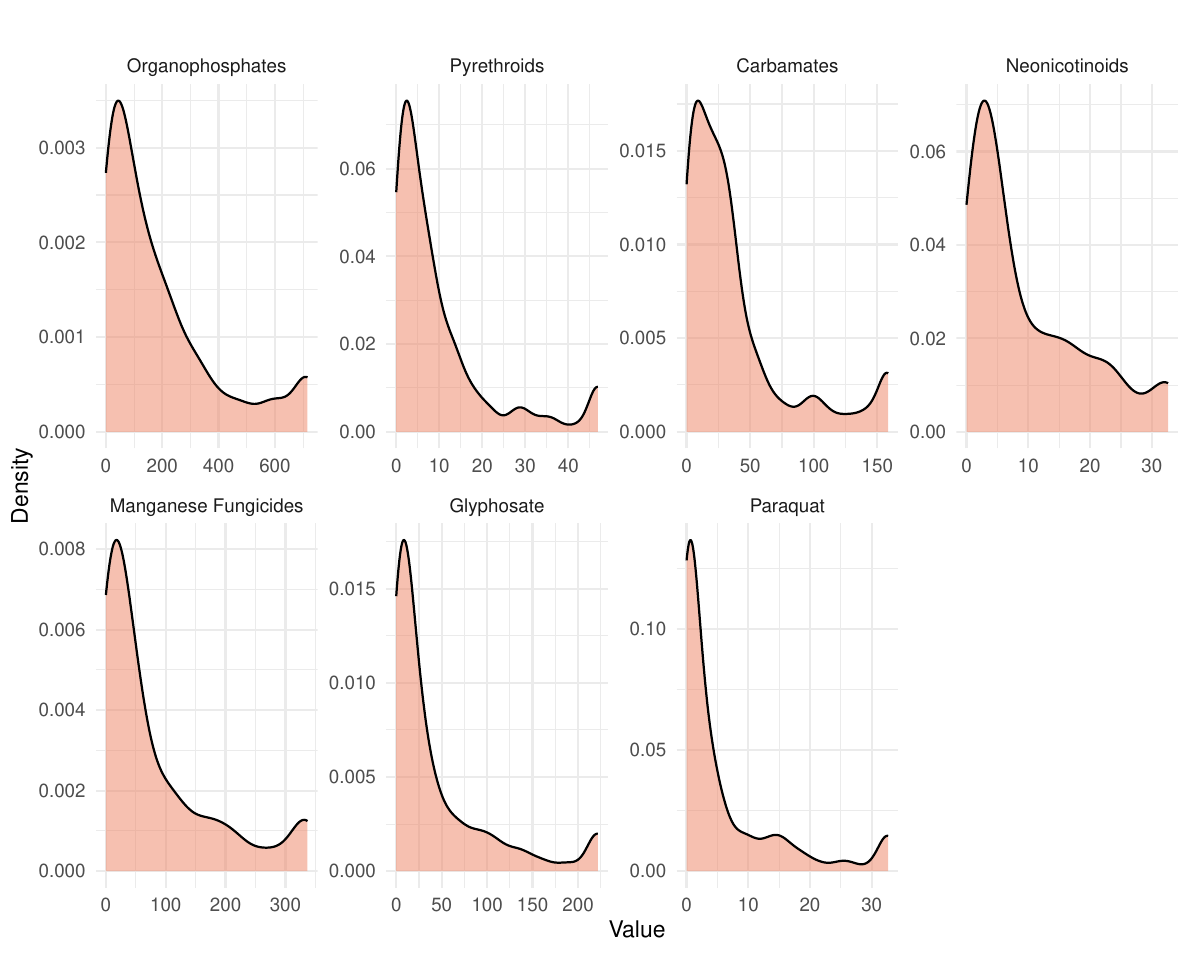}
\label{fig:baselinedis}
\end{subfigure}
\hfill
\begin{subfigure}[t]{0.49\textwidth}
\caption{Spearman correlation matrix.}
\centering
\vspace{-1cm}
\includegraphics[width=1\textwidth]{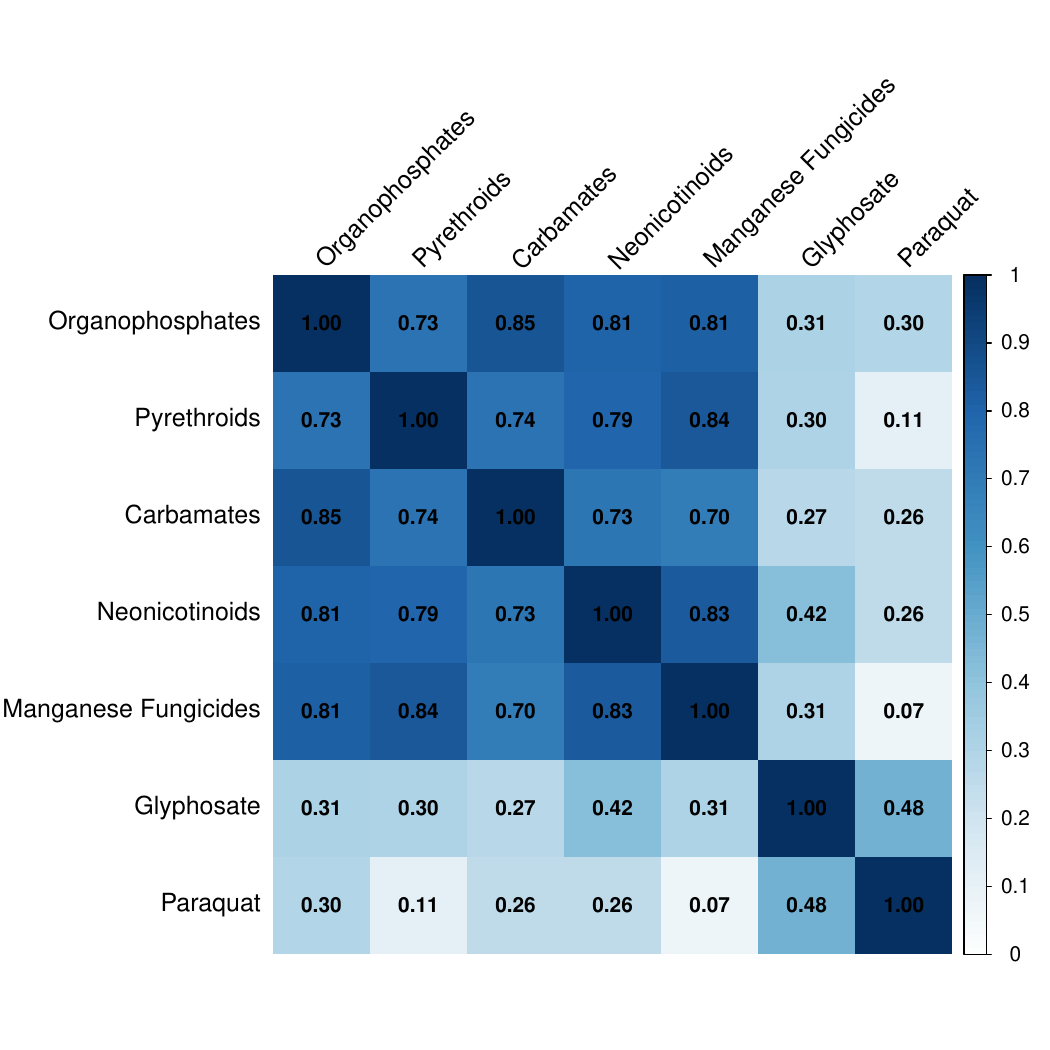}
\label{fig:baselinecorr}

\end{subfigure}
\end{figure}




We consider how to formulate statistical estimands that involve associations between an outcome and one or more continuous mixture components that may be 
(strongly) correlated, 
both 
in the longitudinal setting, $\bar{\mathbf{A}}$, and at a single timepoint, $\mathbf{A}_t$. The applied literature typically considers 
the exposure mixture at a single timepoint and 
three types of causal estimands: 
1) setting the exposure components each to a particular value, 2) limiting the exposure components to be under some policy-relevant threshold, and 3) shifting the exposure components downward either on the additive or multiplicative scale.\cite{smith2023estimating} The first and second type of estimands are not able to be estimated using the nonparametric estimators we discuss herein, as they correspond to parameters that are not pathwise differentiable in the nonparameteric model.\cite{williams2025comment}  This is a technical issue; intuitively, 
because interventions 1 and 2 set many observations to a single value, the intervened-on distribution is nonsmooth, and we can no longer use standard statistical theory for nonparametric model inference.\cite{bickel1993efficient} Only the third estimand that corresponds to shifting the mixture may be nonparametrically estimated;\cite{haneuse2013estimation,diaz2023nonparametric} therefore, we focus on that type of estimand for the remainder of the paper. We also note that choosing the particular shift in the mixture distribution may be an iterative process---one may evaluate the extent to which a 
shift is supported by real data, and if poorly supported, modify the shift to be 
better supported (though we acknowledge that standard uncertainty quantification would not reflect such iteration and that there are alternative approaches to address support issues). This idea has also been discussed by others,\cite{antonelli2024causal} and we consider approaches for doing so in Section \ref{sec:support}.

\subsection{Estimand that quantifies the relationship between a shift in the exposure mixture and the outcome}
\label{sec:mtps}

Estimands that reflect a hypothetical shift of an exposure mixture are commonly considered (e.g.,~\cite{keil2021bayesian,welch2022associations,welch2023racial,mccoy2023semiparametric}), although nearly always in the single timepoint setting.\cite{carone2020pursuit} We briefly summarize 
the single timepoint setting here, then spend more time on the longitudinal setting and situating these estimands within the statistical literature on modified treatment policies.\cite{haneuse2013estimation,young2014identification,diaz2023nonparametric} 

A hypothetical shift of a point-in-time or longitudinal exposure mixture distribution is an example of a \textit{modified treatment policy} (MTP), which is a type of hypothetical intervention on the exposure that is a function of the \textit{natural value of the exposure}. In the single timepoint setting, the natural value of the exposure is just the observed exposure. In the longitudinal setting, the natural value of the exposure is the value the exposure would take at timepoint $t$ if the hypothetical intervention was applied sequentially from baseline and discontinued right before time $t$.
\cite{haneuse2013estimation,young2014identification,diaz2023nonparametric,hoffman2024studying} 
 Importantly, the modified treatment policy may flexibly affect a multi-component exposure, 
 possibly affecting each component differently, and some not at all.

Consider a baseline exposure $\mathbf{A}_0$, 
where we denote a hypothetical intervention 
that is only a function of the natural value of the exposure as $\dd(\mathbf{A}_0)$. Each discrete component 
can be set to a particular value, e.g., $A_{j,0}=a_{j,0}$. 
Continuous components 
can be shifted 
on the additive or multiplicative scale. We can define an additive shift as: $\dd(\mathbf{A}_0) = (A_{1,0} + \gamma_1, A_{2,0} + \gamma_2, ..., A_{7,0} + \gamma_7);$ e.g., 
we could shift each pesticide up by 10 units, $\gamma_i=10$ for each $i \in \{1, .., 7\}$, shown in 
Figure \ref{fig:shifts_all} Panel A. 
 We can define a multiplicative shift as \begin{equation}
 \label{eq:dallbaseline}
     \dd^{\text{all}}(\mathbf{A}_0) = (\delta_1 A_{1,0}, \delta_2 A_{2,0}, ... \delta_7 A_{7,0});
 \end{equation} e.g., we could shift each pesticide down by 20\%, 
 $\delta_i=0.8$ for each $i \in \{1, .., 7\}$, shown in 
 Figure \ref{fig:shifts_all} Panel B. Again, these shifts can vary across components (e.g. a 20\% shift for one and a 10\% shift for another). 
Estimands reflecting shifted exposure values have a similar interpretation to a regression coefficient: 
\textit{a shift in some or all components of $\mathbf{A}$, defined by $\dd$}, \textit{is associated with an X-unit increase in outcome $Y$, holding all other unshifted exposure components (if any) at their observed levels and controlling for covariates.} %

 \begin{figure}[H]
 \caption{Shifts applied to baseline pesticide exposures. (A) Increasing baseline pesticide exposures by 10 (scaled) units. (B) Decreasing baseline pesticide exposures by 20\%}
 \centering
\includegraphics[width=1\textwidth]{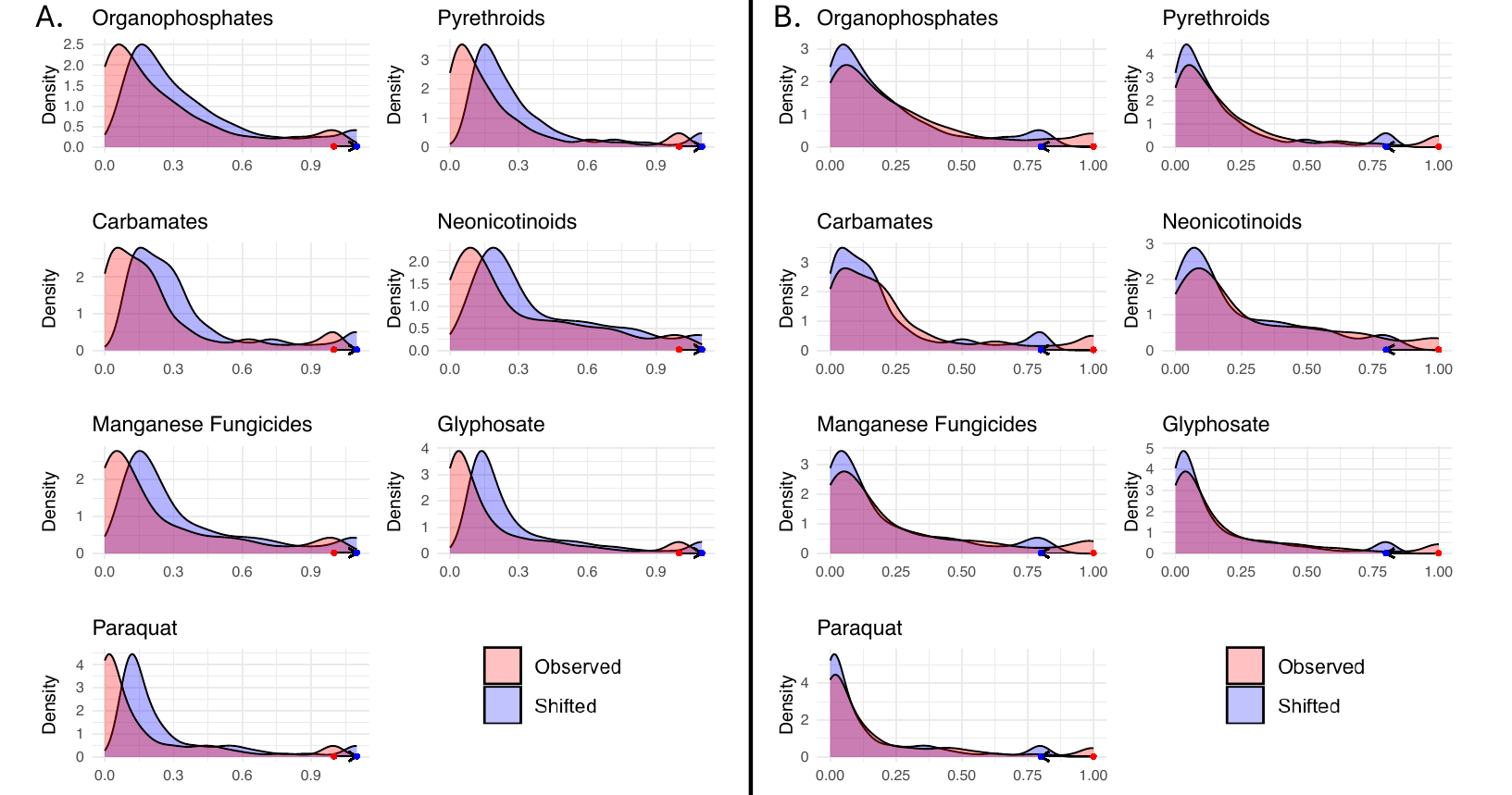}
\label{fig:shifts_all}
\end{figure}

We could also consider a longitudinal setting, where the shifts can vary across mixture components \textit{and} across time. Shifts in time-varying exposure mixtures have rarely been considered in the applied literature. 
We denote the shift applied to $\mathbf{A}$ at each timepoint $t$ as 
 $\mathbf{A}_t^{\dd} = \dd(\mathbf{A}_t(\bar{\mathbf{A}}_{t-1}^{\dd}))
 ,$ 
 where $\mathbf{A}_t(\bar{\mathbf{A}}_{t-1}^{\dd})$ denotes the natural value of $\mathbf{A}$ at time $t$, 
 and the overbar is used to denote the history of $\mathbf{A}$ up to and including the time in the subscript. Consider a simple scenario where $\mathbf{A}$ is a single pesticide measured at two timepoints: $t=0$ and $t=1.$  
 We discussed above that when $t = 0$, 
 an MTP that reduces the natural value of the pesticide by 20\% would yield $\mathbf{A}^\dd_0 = 0.8\times\mathbf{A}_0$. Moving to the second timepoint, $t = 1$, the natural value of exposure at this timepoint is no longer the observed value of the pesticide. Instead, it's the hypothetical value we would have observed in a world where the observed value of the pesticide at baseline had been replaced with its 20\% reduction: 
 $\mathbf{A}_1(\mathbf{A}^\dd_0) = \mathbf{A}_1(0.8\times\mathbf{A}_0)$. Now, we would denote the shift applied to the second timepoint $t=1$ as $\mathbf{A}^\dd_1 = 0.8 \times \mathbf{A}_1(\mathbf{A}^\dd_0) = 0.8 \times \mathbf{A}_1(0.8 \times \mathbf{A}_0)$. 
Importantly, under the identification assumptions given next, the distribution of the natural value of treatment at each time $t$ is equal to the distribution of the observed value of treatment at time $t$, conditional on the observed history. 
For example, we can define a  20\% reduction across each pesticide and at each time as:  \begin{equation}
 \label{eq:alllongitudinal}
     \dd^{\bar{\text{all}}}(\mathbf{A}_t(\bar{\mathbf{A}}_{t-1}^{\dd^{\bar{\text{all}}}})) = 0.8 \times \mathbf{A}_t(\bar{\mathbf{A}}_{t-1}^{\dd^{^{\bar{\text{all}}}}})
 .
 \end{equation}
Similarly, we could define a shift that reduces paraquat levels by 20\% across all five timepoints while leaving the other mixture components unchanged:
\begin{equation}
    \label{eq:a7longitudinal}
\dd^{\bar{A7}}(\mathbf{A}_t(\bar{\mathbf{A}}^{\dd^{\bar{A7}}}_{t-1})) = (A_{1,t}(\bar{\mathbf{A}}^{\dd^{\bar{A7}}}_{t-1}), A_{2,t}(\bar{\mathbf{A}}^{\dd^{\bar{A7}}}_{t-1}), ..., 0.8 \times A_{7,t}(\bar{\mathbf{A}}^{\dd^{\bar{A7}}}_{t-1})).
\end{equation}

For any shift defined by $\dd$, we can consider the following causal estimand: 
           \begin{equation}
           \label{eq:causalest}
               \E(Y(\bar{\mathbf{A}}^{\dd}) - Y),
           \end{equation}
where $Y(\bar{\mathbf{A}^{\dd}})$ is the counterfactual outcome had the shift $\dd$ been applied to the longitudinal natural values of the exposure mixture through the end of follow-up ($\bar{\mathbf{A}}^{\dd}$) and $Y$ is the observed outcome. 
Using the multiplicative shift in Eq  \ref{eq:alllongitudinal}, we would interpret the above estimand as the average difference in risk of hypertension had pesticide exposure levels been 20\% less than what they actually were across all 7 classes of pesticides and across all 5 exposure times versus observed hypertension. 
 
 Under the following identification assumptions we can equate the causal estimand in Eq \ref{eq:causalest} with \textit{statistical estimand} in Eq \ref{eq:statest}; the statistical estimand is only a function of observed data and thus, estimable.\cite{diaz2023nonparametric,hoffman2024studying} The strong sequential randomization assumption assumes that all confounders of the exposure at each time and all future exposures, time-varying covariates, and outcomes are measured and conditioned on. Positivity assumes that if it is possible to find an observation with some exposure and history at time $t$ (where the relevant history is the set of variables needed to condition on to satisfy the strong sequential randomization assumption), then it is possible to find an observation with the same history at time $t$ and shifted exposure value. The statistical estimand is 
 \begin{equation}
  \E[...\E[\E(Y \mid \mathbf{A}_4=\dd(\mathbf{A}_4(\bar{\mathbf{A}}_3^{\dd})), H_4) \mid \mathbf{A}_3=\dd(\mathbf{A}_3(\bar{\mathbf{A}}_2^{\dd})), H_3]  ... \mid \mathbf{A}_0=\dd(\mathbf{A}_0), H_0],
     \label{eq:statest}
 \end{equation}
where $H_t$ are the variables needed to condition on to satisfy the strong sequential randomization assumption, stated above. We further consider the positivity assumption in Section \ref{sec:support}.  

Code to visualize the observed and shifted distributions is available: \url{https://github.com/CI-NYC/CHAMACOS/blob/main/scripts/0_shift_plots.R}.

\section{Using our shifted exposure mixture estimand to assess interactions}
The third common scientific question about mixtures listed in Section \ref{sec:intro} is the extent to which their components exhibit interactive effects. An interaction between two binary exposures, $A_1$ and $A_2$, on the causal additive scale is defined as\cite{vanderweele2009distinction}
\begin{equation}\label{eq:interaction}
    \E[Y(1,1)] - \E[Y(0,1)] \neq \E[Y(1,0)] - E[Y(0,0)],
\end{equation}
where $Y(a_1,a_2)$ is the counterfactual value of $Y$ setting $A_1=a_1,A_2=a_2$. 
Here, we use McCoy et al.'s extension to a nonparametric definition of interaction of continuous mixture components.\cite{mccoy2023semiparametric}   

We can situate this prior work in the longitudinal setting with an example: 
examining the interaction between longitudinal glyphosate exposure ($\bar{A}_6$) and longitudinal paraquat exposure ($\bar{A}_7$) on risk of hypertension. 
Using the work of \cite{mccoy2023semiparametric}, the null hypothesis of no interaction can be defined as: 
\begin{equation*}
    H_0: \E[Y(\bar{\mathbf{A}}^{\dd^{\bar{A6},\bar{A7}}})] - \E[Y(\bar{\mathbf{A}}^{\dd^{\bar{A7}}})] - \E[Y(\bar{\mathbf{A}}^{\dd^{\bar{A6}}})] + \E[Y] =0.
\end{equation*}
We clarify that other components of our mixture may also interact---that is fine!---we focus on the potential interaction of these two components here.

Code to nonparametrically test the null hypothesis of no interaction of components of the exposure mixture---both in the single timepoint and longitudinal settings---is available: \url{https://github.com/CI-NYC/CHAMACOS/blob/main/scripts/5_interaction_hypothesis_testing.R}. See Section \ref{sec:estimators} for estimator details and \ref{sec:supp:hypothesis} for statistically testing this hypothesis.

\section{Question: How can we evaluate the extent to which shifts in the exposure mixture, $\dd$, have support in our data?} 
\label{sec:support} 
In Section \ref{sec:mtps}, we showed how one could flexibly define 
shifts across one or more members of the exposure mixture and across one or more timepoints. However, it is well-known that not all shifts will be well-supported by the observed data.\cite{snowden2015framing,zigler2021invited,keil2021keil,carone2020pursuit,antonelli2024causal} If the shift we first consider is not well-supported, meaning that it is based on regions where there are no real-world data, 
is it still of interest? Many would answer no, as such a shift would not reflect the real world.\cite{zigler2021invited} 

\subsection{Positivity}
T
o be ``in support" of the observed data 
means satisfying the positivity assumption defined 
in Section \ref{sec:mtps}. Informally, positivity means that 
we do not condition on zero-probability events. For the estimands we consider, 
in the single timepoint case, it means that for every exposure-covariate combination observed in our data, there is a nonzero probability of observing the shifted exposure value conditional on the covariates; and in the longitudinal case, it means 
that for every observed value of the exposure mixture and history at time $t$, the exposure mixture's shifted value must also be in the support of the data. We can formalize this assumption 
as: if $(\mathbf{a}_t, h_t)\in supp\{\mathbf{A}_t, H_t\},$ then $(\dd(\mathbf{a}_t, h_t), h_t)\in supp\{\mathbf{A}_t, H_t\}$ for each $t\in \{0, .., T\}$,\cite{diaz2023nonparametric} where $H_t$ denotes the history at time $t$.\cite{diaz2023nonparametric,hoffman2024studying}

Because 
positivity is a function only of observed data---not counterfactual data---
we can evaluate it 
and the extent to which there may be near-positivity violations, which occur when the conditional exposure probabilities are not strictly zero but are close to zero.\cite{petersen2012diagnosing} In the mixture setting of multiple continuous exposures and numerous covariates, satisfying positivity is exceptionally challenging,\cite{bao2025addressing} even more so when considering longitudinal estimands. Bao and Schomaker propose examining multidimensional density ratios to assess positivity, which we discuss in Section \ref{sec:bao}.\cite{bao2025addressing} However, here, we explore how one can use an alternative, cruder measure to evaluate extrapolation---the convex hull. 

\subsection{Using a convex hull to explore extrapolation}
\label{sec:convexhull}
A \textit{convex hull} of observed data across $n$ variables is an n-dimensional polygon that is the smallest convex set that contains those data points.
\cite{king2006dangers} 
In the exposure mixture setting, it is ``the smallest polygon 
where exposure values are actually observed"
.\cite{antonelli2024causal} 
The convex hull is the border between observed or interpolated data and---beyond the border---extrapolation. Thus, regions inside the convex hull may have poor or no support in the observed data, but such 
interpolation is typically more palatable than extrapolation.\cite{king2006dangers} 
 Use of the convex hull to assess support for environmental mixtures was proposed and explored with both simulated and real-world data recently.\cite{antonelli2024causal}

We aim to lower the barrier for applied researchers to use the convex hull in their own work by providing open-source software (as a Julia module for computational efficiency) that: 1) takes as input the matrix of observed exposure mixture values and outputs the convex hull, and 2) takes as input the convex hull and the matrix of exposure mixture values under the desired shift and outputs a matrix of shifted exposure values that are within the convex hull (where shifted values that would lie outside the hull are instead shifted to the closest point on the hull). The software is available: \url{https://github.com/nt-williams/MTPConvexHull.jl}

We illustrate this by using the marginal observed pesticide exposure values to define a 7-dimensional convex hull at each time $t$ and assess the extent to which various shifts, $\dd$, would lie inside or outside this polygon. Although there are other options for defining the convex hull of longitudinal exposure mixtures, such as defining a single convex hull 
by considering observed values at all timepoints, we proceed with this timepoint-specific 
approach.
Because the convex hull of marginal exposure values represents a less restrictive space than the convex hull of exposure values conditional on covariates,
\cite{antonelli2024causal} using the marginal values would identify some but not all of the positivity issues. Note that although we calculate the 7-dimensional convex hull, visualizing a 7-dimensional object is near impossible. Figure \ref{fig:3dsupport} shows the convex hull of three 
pesticides for simplicity. 

\begin{figure}[H] 
\captionsetup{justification=raggedright,singlelinecheck=false}
  \caption{Convex hull of 3 of the 7 pesticides at baseline: pyrethroids (X-axis), neonicotinoids (Z-axis), and manganese-containing fungicides (Y-axis).}
\centering
\includegraphics[width=0.6\textwidth]{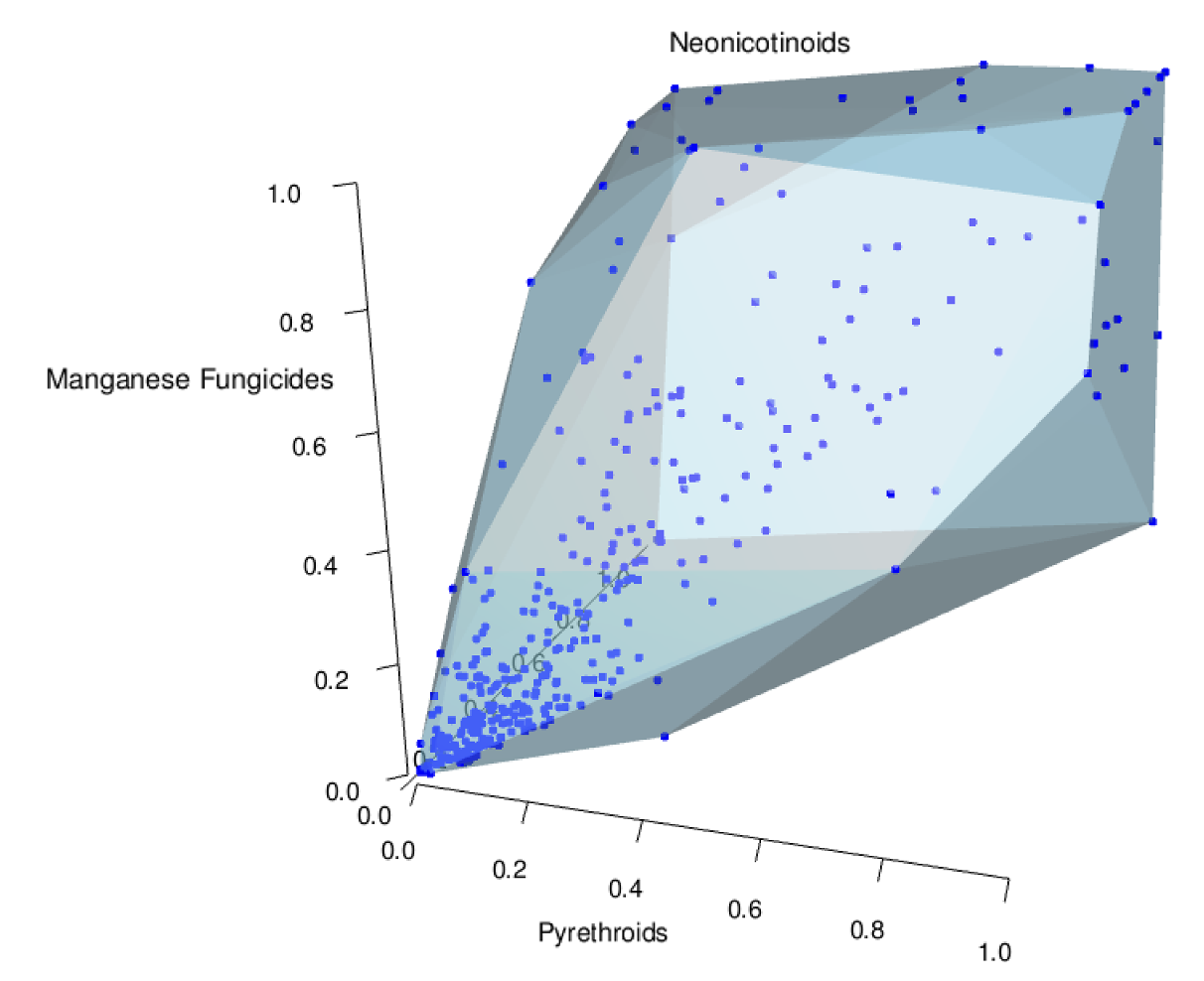}
\label{fig:3dsupport}
\end{figure}

Antonelli and Zigler\cite{antonelli2024causal} consider one metric 
of extrapolation, denoted R, 
that represents the proportion of the shifted distance that lies outside the convex hull ($\frac{\text{distance[closest point on the convex hull to the desired shifted point, shifted point]}}{\text{distance[observed point, shifted point]}}$). We consider both R (a relative measure) and its numerator (an absolute measure). 
We add the absolute measure, because many individuals have low levels of exposure (Figure \ref{fig:baselinedis}), and a 20\% reduction of a number close to zero would be another number close to zero, making the denominator close to zero. We may not care about a large R value in such cases, as the absolute value of the extrapolation would also be close to zero. The degree of extrapolation that someone is comfortable with is likely subjective and problem-specific. Antonelli and Zigler set 
R$<0.1$ to represent a minimal amount of extrapolation.\cite{antonelli2024causal} We consider this same cut-off and use $\le0.05$ standardized units (exposure levels standardized to the [0,1] scale) for the absolute measure to represent a minimal amount of extrapolation, which translates to about one-quarter of a standard deviation of each pesticide.  


\subsection{A warning about assessing extrapolation based only on the correlation matrix; how one can 
use the multidimensional convex hull of the exposure mixture 
instead}
Consider a 20\% reduction of glyphosate and paraquat at baseline: $\dd^{A6,A7}(\mathbf{A}_0)$. 
In our dataset, these two components are not very correlated with each other or other pesticide classes (Figure \ref{fig:baselinecorr}), so such a shift 
seems unlikely to result in extrapolation. 
However, examining the convex hull, 
\textit{over 16\% of shifted exposure values lie outside the convex hull. Most of these---over 14\% of all shifted exposure values---have an R value $>0.1$.} 
 But the absolute amount of extrapolation for many of these observations is low; 
 2.7\% of observations overall 
 would extrapolate beyond 0.05 standardized units. 

We simulate two joint exposures that more dramatically illustrate the potential benefit of the convex hull over the Spearman correlation matrix to assess extrapolation \\ \noindent(\url{https://gist.github.com/nt-williams/06c1acac46db0b32c82cecc51fc0d62b}). Because Spearman correlation only measures linear dependence between mixture components (may be nonmonotonic), complex nonlinear relationships will not be captured. In our simple simulation, the two mixture components are uncorrelated; however, a 20\% reduction results in 43\% of observations extrapolated beyond the convex hull with R$>0.1$ and 23\% extrapolated beyond 0.05 standardized units.

Consider a shift that may be more ill-advised. 
Say we 
would like to estimate the association between a 20\% reduction in neonicotinoid exoposure at baseline and hypertension risk, i.e., a shift of $\dd^{A4}(\mathbf{A}_0) = (A_{1,0}, A_{2,0}, A_{3,0}, 0.8 \times A_{4,0}, A_{5,0}, A_{6,0}, A_{7,0})$. This would be ill-advised based on Figure \ref{fig:baselinecorr}, because of strong correlation between neonicotinoid levels and $A_{1,0}, A_{2,0}, A_{3,0}, A_{5,0}$. 
Indeed, we find that such a 20\% reduction would result in \textit{over 36\% of the shifted exposure values lying outside of the convex hull}, nearly all of which 
have an R value$>0.1$, and 45\% of which extrapolate $>$0.05 standardized units. 
\textbf{Extrapolation is not the only issue with this particular shift.} When we consider the set of supported shifted values reflecting this 20\% reduction in neonicotinoids---``supported shifted" meaning that the shifts are within the convex hull or, if outside of it, 
shifted to 
the closest point on the convex hull---
the levels of non-neonicotinoid pesticides are, on average, \textit{greater} than their original, non-shifted exposure levels. In other words, a substantial subset of points on or within the convex hull that reflect a 20\% \textit{reduction} in neonicotinoid levels require \textit{increased} levels of other pesticides. 
For example, 19\% of supported shifted observations have higher glyphosate exposure values than observed and 9-12\% have higher values of other pesticide classes. 
It may be counterintuitive to think that a relatively small 20\% reduction in one pesticide would necessitate upward shifts in other pesticides in order to stay on or within the convex hull of support, but this may reflect  substitution of one class of pesticide for another. 

Code to output the convex hull, evaluate extrapolation under the two example shifts we consider, and output the closest point on the convex hull to shifts that would fall outside of it is available: \url{https://github.com/CI-NYC/CHAMACOS/blob/main/scripts/1_convex_hull_rep_section4.R}.

\section{Question: How do we define mixture shifts that are better supported by our data? }
\label{sec:shiftsinhull}
Others have suggested several approaches for defining better-supported mixture shifts, including: 1) focusing on reductions on the multiplicative scale, 
because of better support down to the lower bound of zero;\cite{keil2021bayesian} 2) examining smaller shifts;\cite{keil2021bayesian,mccoy2023semiparametric,keil2024considerations} 3) avoiding shifts of certain variables or variable combinations that are highly correlated with other, unshifted, mixture components;\cite{snowden2015framing,bobb2015bayesian} and 4) redefining the shift such that shifted observations that would rely on extrapolation are not shifted and instead, left as observed.\cite{diaz2023nonparametric} Approaches 1-3 have been discussed frequently in previous literature. 
We implement these suggestions by considering small, multiplicative reductions of 20\% of variable combinations that are likely to be well-supported. Below, we formalize and illustrate the fourth approach. 

First, though, we note that an alternative to redefining the shift is to redefine the estimand to average over a different population. Antonelli and Zigler\cite{antonelli2024causal} give two suggestions: 1) downweight shifted observations that rely more heavily on extrapolation and upweight those that are better supported; and 2) average over the population with shifted values within the the convex hull. We discuss a similar idea in Section \ref{sec:localsupp}, redefining the estimand to average over a subset of the population defined by certain exposure values---a generalization of the average treatment effect on the treated.\cite{susmann2024local}

To illustrate the fourth approach of redefining the shift, 
we can redefine 
$\dd^{\bar{A6}, \bar{A7}}(\mathbf{A}_t(\mathbf{A}^{\dd^{\bar{A6}, \bar{A7}}}_{t-1}))$ as 
shifting the longitudinal mixture values only if the shifted 7-dimensional point stays within the convex hull, but leaving at its observed 
value otherwise. 
We might also be comfortable with some extrapolation, especially after realizing that the vast majority of the shifted exposure values outside the convex hull are not far outside of it. If so, we could redefine $\dd^{\bar{A6}, \bar{A7}}(\mathbf{A}_t(\mathbf{A}^{\dd^{\bar{A6}, \bar{A7}}}_{t-1}))$:
\begin{equation}
\label{eq:a6a7longitudinalhull}
\dd^{\bar{A6}, \bar{A7}}(\mathbf{A}_t(\mathbf{A}^{\dd^{\bar{A6}, \bar{A7}}}_{t-1})) = \begin{cases}
 (A_{1,t}(\mathbf{A}^{\dd^{\bar{A6}, \bar{A7}}}_{t-1}), ..., 0.8 \times A_{6,t}(\mathbf{A}^{\dd^{\bar{A6}, \bar{A7}}}_{t-1}), 0.8 \times A_{7,t}(\mathbf{A}^{\dd^{\bar{A6}, \bar{A7}}}_{t-1})) &\text{ if }\\
 \text{distance}[\mathbf{A}_t^{\text{convex hull}}, (A_{1,t}, ..., 0.8 \times A_{6,t}, 0.8 \times A_{7,t})] \le 0.05 \\
 \mathbf{A}_{t}(\mathbf{A}^{\dd^{\bar{A6}, \bar{A7}}}_{t-1})  & \text{ otherwise,}
\end{cases}
\end{equation}
where $\mathbf{A}_t^{\text{convex hull}}$ represents the point on the convex hull closest to the desired shift, and 
where we 
apply the shift if the amount of extrapolation is $\le0.05$ standardized units. There are 
infinitely many options; alternatives could require some minimum density of the marginal joint exposure distribution, consider various thresholds for extrapolation, or incorporate positivity directly.

Code to define the shifts we consider here is available: \url{https://github.com/CI-NYC/CHAMACOS/blob/main/scripts/1c_convex_hull_last_2_shift.R}.


\section{Question: How can we estimate point-in-time and longitudinal mixture shifts 
using flexible and computationally scalable nonparametric estimators?}
\label{sec:estimators}
 We use software\cite{lmtp2023} that implements two nonparametric estimators in single timepoint and longitudinal settings: targeted minimum loss-based estimation (TMLE) \cite{diaz2023nonparametric,van2012targeted} and sequentially doubly robust (SDR) estimation,\cite{diaz2023nonparametric,luedtke2017sequential,rotnitzky2017multiply} the statistical properties of which have been discussed previously.\cite{diaz2023nonparametric, hoffman2024studying} 
Compared to a parametric estimator, TMLE and SDR have two key advantages: 1) they are robust to misspecification of certain combinations of models that comprise the estimator, 
a property called \textit{double robustness} (they have slightly different robustness properties, explained in \ref{sec:estimatorsupp}); and 2) allow for machine learning to fit these nuisance parameter models while retaining theoretically valid inference---called \textit{rate double robustness}.\cite{kennedy2024semiparametric} As it is implausible to correctly parametrically specify the relationships we consider in our motivating example, we use an ensemble of machine learning algorithms, called the \textit{super learner} (also referred to as stacking) algorithm,\cite{van2007super} which combines candidate models through a weighted convex combination (i.e., weights are non-negative and sum to one) using cross-validation. 

The \textit{lmtp}\cite{lmtp2023} package we use for estimation uses cross-fitting to obtain out-of-sample estimate of nuisance parameters, which reduces risk of over-fitting.\cite{zheng2011cross} For particularly large datasets (e.g., millions of observations), one can use a more computationally efficient version (available \url{https://github.com/nt-williams/lmtp/tree/mlr3superlearner})
\cite{mlr3superlearner} based on the \textit{mlr3}\cite{mlr3} package. 

Code to estimate single timepoint and longitudinal shifts using these nonparametric estimators is available: \url{https://github.com/CI-NYC/CHAMACOS/blob/main/scripts/3a_analysis_single_timepoint.R}, \url{https://github.com/CI-NYC/CHAMACOS/blob/main/scripts/2_analysis_longitudinal.R}.

\section{Real-world application}
We put the above discussion into practice and estimate the relationship between several longitudinal 
20\% exposure mixture reductions on the risk of hypertension. We include a simpler, single timepoint analysis in \ref{sec:supp:analysis}. 
The CHAMACOS cohort and their longitudinal exposure to the pesticide mixture were introduced in the Introduction and discussed in \ref{sec:supp:cohort} and \ref{sec:supp:measures}. 

\subsection{Statistical analysis: longitudinal data application}
In the longitudinal setting, we have observed data $O=(\mathbf{L_0}, \mathbf{A_0}, C_1, $ $Y_1, ..., \mathbf{L_4}, \mathbf{A_4}, C_5, Y_5)$
where $\mathbf{L_{t}}$ denotes the vector of covariates at time $t$ (including baseline covariates at $t=0$), 
$C_{t}$ denotes censoring by time $t$, and $\mathbf{A_{t}}$ and $Y_{t}$ were defined in the Introduction. 
Time-invariant baseline covariates that we adjusted for included: education level, maternal age, maternal age at delivery of the child, born in the US, history of having a high blood pressure diagnosis or diabetes diagnosis, and cohort wave of enrollment. 
Time-varying covariates that we adjusted for included: marital status, poverty level, presence of agricultural workers in the household since the previous visit, and work status since the previous visit. 
We are interested in the causal estimand: $$\lambda^* = \E(Y(\bar{\mathbf{A}}^{\dd^*}, \bar{C}=\bar{1}) - Y(\bar{C}=\bar{1}) ),$$ for $\dd^* \in \{\dd^{\bar{\text{all}}}, \dd^{\bar{\text{first}5}}, \dd^{\bar{A6}, \bar{A7}}, \dd^{\bar{A7}}\}$; these four shifts were chosen because they do not rely much, if at all, on extrapolation. $\dd^{\bar{\text{all}}}$ is given in Eq\ref{eq:alllongitudinal}, $\dd^{\bar{\text{first}5}}$ is defined below, $\dd^{\bar{A6}, \bar{A7}}$ is given in Eq\ref{eq:a6a7longitudinalhull}, and $\dd^{\bar{A7}}$ is given in Eq\ref{eq:a7longitudinal}. 
In words, $\lambda^{\bar{\text{first}5}}$ is the expected difference in risk of developing chronic hypertension by the final visit had everyone's exposure levels of organophosphates, pyrethroids, carbamates, neonicotinoids, and manganese fungicides been reduced by 20\% at each of the five visits (leaving their levels of glyphosphate and paraquat as observed), as compared to observed hypertension risk. $\lambda^{\bar{\text{all}}}$, $\lambda^{\bar{A6},\bar{A7}}$ and $\lambda^{\bar{A7}}$ are defined similarly. For $\dd^{\bar{\text{all}}}$, $\dd^{\bar{\text{first}5}}$, $\dd^{\bar{A6}, \bar{A7}}$, $\dd^{\bar{A7}}$, the percent of observations $\ge$0.05 standardized units outside the convex hull are 0\%, 0.2\%, 0.7\%, and 0.2\%, respectively. 
In our definition of these shifts, we leave the exposure levels as observed if the shifted values would extrapolate beyond this amount (e.g., Eq \ref{eq:a6a7longitudinalhull}). 

We used TMLE with 20 cross-fit folds to estimate these longitudinal effects.\cite{diaz2023nonparametric} This estimator adjusts for baseline and time-varying covariates in the exposure, censoring, and outcome models. 
We estimated outcome nuisance parameters using a stacked ensemble\cite{van2007super} of: an intercept-only model, a main-effects generalized linear model, LASSO,\cite{tibshirani1996regression} multivariate adaptive regression splines\cite{friedman1991multivariate}, random forests\cite{breiman2001random}, and several implementations of light gradient boosting\cite{ke2017lightgbm}. 

Commented, step-by-step code for the longitudinal analysis is available \url{https://github.com/CI-NYC/CHAMACOS/blob/main/scripts/2_analysis_longitudinal.R}

\subsection{Results: longitudinal data application}
Figure \ref{fig:longredall} shows point estimates and 95\% confidence intervals of each $\hat \lambda^*$. 
We estimate that longitudinally reducing all pesticide classes by 20\% would reduce risk of hypertension by the last follow-up visit by 4.4 percentage points (95\% CI: -0.2, 11.0). 
Longitudinally reducing just the first 5 pesticide classes, last 2 pesticide classes, or paraquat only  
were estimated to reduce risk of hypertension by similar or lesser amounts (risk reductions of 4.9 (95\% CI: -0.6, 10.3), 1.1 (95\% CI: -4.0, 6.1), and 1.0 percentage points (95\% CI: -1.1, 2.9), respectively). 

This suggests that---if our identification assumptions hold---longitudinal 20\% reductions to this set of seven pesticide classes over 10 years is estimated to reduce risk of 
hypertension by the final follow-up point by over 4 percentage points, when the mothers are 48 years old, on average. Our results suggest that exposure to the first five pesticide classes may contribute more to this risk than exposure to glyphosates and paraquats.

\begin{figure}[H] 
\captionsetup{justification=raggedright,singlelinecheck=false}
  \caption{Effect estimates of longitudinal 20\% reduction across pesticide classes and across timepoints, which are follow-up waves.}
\centering
\includegraphics[width=1\textwidth]{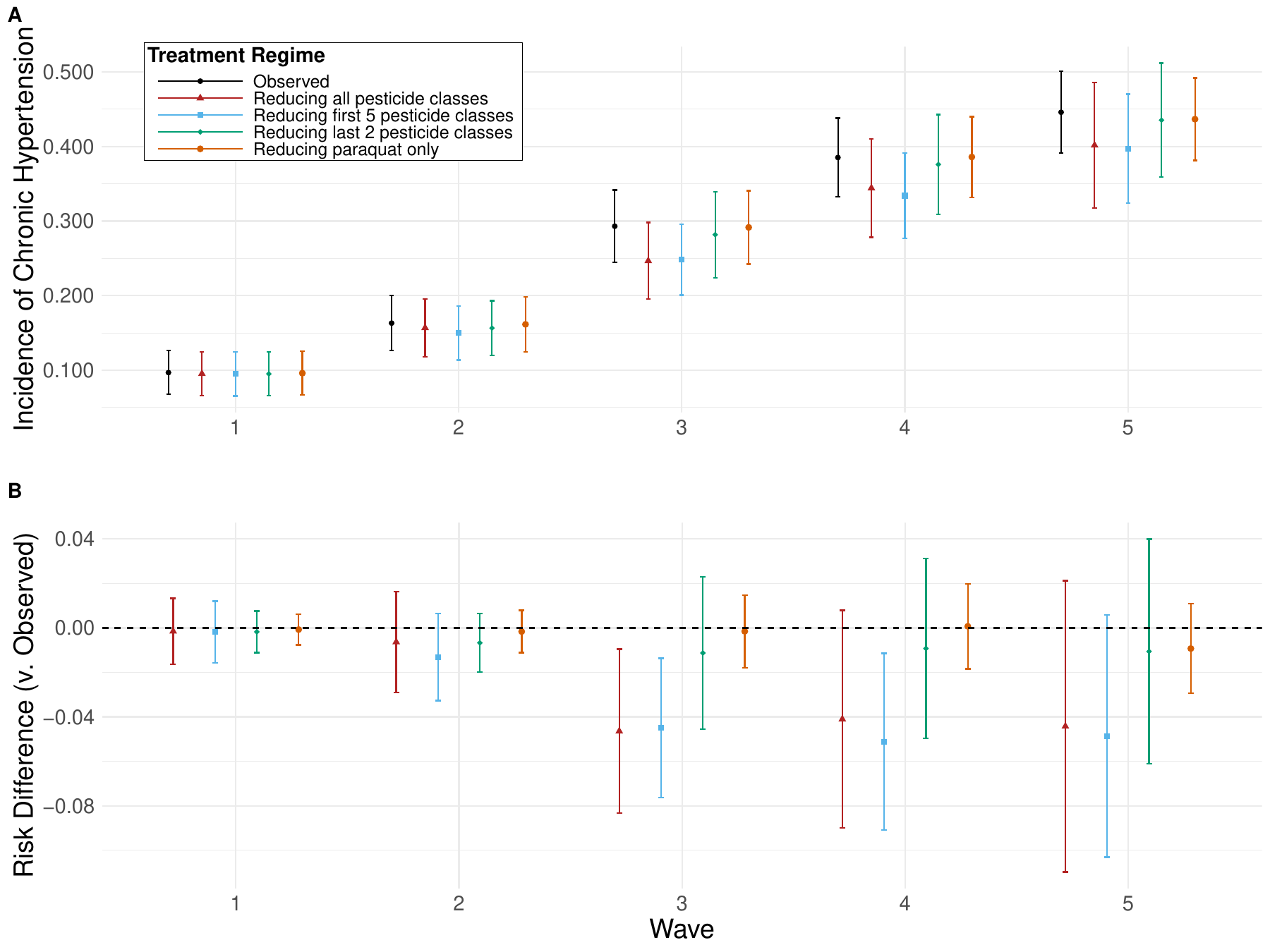}
\label{fig:longredall}
\end{figure}

\section{Discussion}
There has been limited work on how to quantify and estimate relationships between environmental mixtures and outcomes in the longitudinal setting when exposure mixtures and outcome are time-varying, including limited nonparametric estimator options that would improve the robustness of results. 
Assessing the extent to which there is data support for these longitudinal estimates is also critical for improving robustness---results are more robust when they are based on real data instead of on extrapolation---but open-source, computationally scalable tools are lacking. In this review and tutorial, we integrate literature on longitudinal modified treatment policies and their nonparametric estimation using ensembles of machine learning algorithms, and support extrapolation assessment via a multi-dimensional, longitudinal convex hull, alongside step-by-step R code (\url{github.com/CI-NYC/CHAMACOS}) and Julia software (\url{https://github.com/nt-williams/MTPConvexHull.jl}) to lower the barriers for applied researchers to use these methods. 

Much additional work remains to improve the robustness of estimating the effects of exposure mixtures. In terms of evaluating support, future work could examine: a) how to best define a convex hull of longitudinal exposure data---evaluating the tradeoffs between defining a separate convex hull at each timepoint versus a single, convex hull that encompasses all timepoints; b) how to formulate a convex hull that is a 
function of both the exposure mixture and relevant covariates, thereby more closely reflecting the positivity assumption; and c) extending such a formulation to natural values of treatment in the longitudinal setting, thereby reflecting the positivity assumption specific to modified treatment policies. In terms of formulating and estimating estimands of longitudinal exposure mixtures that are better supported---or at least more transparent about the extent of support, future work can 1) evaluate different approaches for decomposing the estimand into: the portion of the estimand that is supported, and the portion that relies on extrapolation, possibly further decomposing by degree of extrapolation; and 2) evaluate different approaches for estimating such decomposed estimands, discussing tradeoffs. We are currently pursuing such work.  
There is also a need for future work to define causal estimands and develop estimators that accommodate longitudinal estimation of mixtures in other study designs, like case-control designs, and to accommodate interference or spatial dependence, which are 
especially relevant to environmental mixtures.

\newpage
\section*{Figure captions}
\begin{figure}[H] 
\captionsetup{justification=raggedright,singlelinecheck=false}
  \caption{Example of extrapolation (2-dimensional red points in the gray shaded region) outside of the convex hull (defined in Section \ref{sec:convexhull} black dashed polygon) when decreasing neonicotinoid values by 50\% while keeping manganese-containing fungicide values as observed. If we considered neonicotinoid as a single pesticide exposure and applied a reductive 50\% shift, we would not extrapolate beyond the bounds of the observed \textit{neonicotinoid} values, as there is support between scaled values of 0 and 1. However, when considering neonicotinoid and manganese-containing fungicide values \textit{jointly} and applying a reductive 50\% shift on the former alone, we find numerous 2-dimensional data points that would extrapolate beyond the bounds of the observed data. In this simple example, we find that there is little support for high-levels of manganese-containing fungicide exposure and simultaneously low-levels of neonicotinoid exposure. Data source: CHAMACOS Maternal Cognition Study}
\label{fig:extrapolation}
\end{figure}

\noindent \begin{figure}[H] 
\caption{Each pesticide class at baseline, truncated at the 95\% percentile.}
\begin{subfigure}[t]{0.49\textwidth}
      \caption{Distribution.}
\label{fig:baselinedis}
\end{subfigure}
\hfill
\begin{subfigure}[t]{0.49\textwidth}
\caption{Spearman correlation matrix.}
\label{fig:baselinecorr}

\end{subfigure}
\end{figure}

 \begin{figure}[H]
 \caption{Shifts applied to baseline pesticide exposures. (A) Increasing baseline pesticide exposures by 10 (scaled) units. (B) Decreasing baseline pesticide exposures by 20\%}
 \centering
\label{fig:shifts_all}
\end{figure}

\begin{figure}[H] 
\captionsetup{justification=raggedright,singlelinecheck=false}
  \caption{Convex hull of 3 of the 7 pesticides at baseline: pyrethroids (X-axis), neonicotinoids (Z-axis), and manganese-containing fungicides (Y-axis).}
\centering
\label{fig:3dsupport}
\end{figure}

\begin{figure}[H] 
\captionsetup{justification=raggedright,singlelinecheck=false}
  \caption{Effect estimates of longitudinal 20\% reduction across pesticide classes and across timepoints, which are follow-up waves.}
\centering
\label{fig:longredall}
\end{figure}

\clearpage
\setcounter{page}{1}

\renewcommand{\arraystretch}{1}
\captionsetup[table]{name=Table} 
\captionsetup[figure]{name=Figure} 
\renewcommand{\thetable}{e\arabic{table}}
\renewcommand{\thefigure}{e\arabic{figure}}
\renewcommand{\theequation}{e\arabic{equation}}
\setcounter{table}{0} 
\setcounter{figure}{0}
\setcounter{equation}{0}
\setcounter{page}{1}
\renewcommand{\thepage}{Supplementary Digital Content \arabic{page}}
\setcounter{section}{0}
\renewcommand{\thesection}{e\arabic{section}}

\clearpage
\raggedright
\appendix


\begin{center}
\LARGE{Supplementary Digital Content for} \\ 
    \vspace{0.5cm} 
    \LARGE{Everything all at once: On choosing an estimand for multi-component environmental exposures} \\ 
    \vspace{0.5cm}  
\end{center}

\raggedright

\renewcommand{\thesection}{e\arabic{section}}
\renewcommand{\thesubsection}{\thesection.\arabic{subsection}}

\setcounter{section}{0} 
\section{Additional detail on the motivating example using the CHAMACOS cohort}

\subsection{Cohort}\label{sec:supp:cohort}
Data come from the Center for the Health Assessment of Mothers and Children of Salinas (CHAMACOS) Maternal Cognition Study. Study activities were approved by the University of California, Berkeley, Office for the Protection of Human Subjects (Protocol ID 2021-02-14055) and written informed consent was obtained from all participants prior to participation. This study builds on a longitudinal birth cohort of low-income, Mexican American families in the Salinas Valley, a California agricultural community.\cite{eskenazi2004association} Initial enrollment occurred in two waves. In 1999-2000, pregnant women attending prenatal care visits at one of six local clinics were screened for eligibility ($\ge$18 years old; $\le$20 weeks gestation; English- or Spanish-speaking; and eligible for California’s Medicaid program, MediCal) and 1,130 women met eligibility criteria. Of these, 597 unique women enrolled into a prospective pregnancy cohort and 525 unique women continued participating with their liveborn index child(ren) at or after delivery. When index children turned 9, 348 unique women were still engaged in the study.  
In 2009-2011, to refresh the cohort, additional mother-child dyads were recruited through primary care clinics, schools, churches, community service agencies, and by referral from existing participants. Women were eligible to enroll in CHAM2 if they were English- or Spanish-speaking and had a 9-year-old child for whom they had received prenatal care in the Salinas Valley. In addition, women needed to have qualified for MediCal during pregnancy and to have been at least 18 when their index child was born. At this point, 287 unique new women enrolled in the cohort, for a total of 635 participants. From that point, families proceeded to complete study roughly every two years between child ages 9 and 18, and to participate in COVID-specific data collection from 2020-2021. 
The CHAMACOS cohort has been supported by the EPA (RD83451301), NIH-NIEHS (P01ES009605, R01ES017054, R01ES021369, R01ES026994, R24ES028529, UH3ES030631, U24ES028529-06), NIH-NIDA (R01DA035300), and NIH-NIA (R01AG069090).

This specific tutorial focuses on maternal participants with available pesticide exposure measures, blood pressure outcomes, and covariate information starting from the 2010-2013 visit (when index children were roughly age 10.5 years). 621 unique women attended any study visit between the 2010-2013 visit to the 2022-2024 visit. We excluded patients with baseline cancer, patients who never had sufficient exposure measurements (i.e. we excluded patients without at least one instance in which exposures were measured for at least 1 year due to insufficient data), and patients who experienced the outcome before the first exposure measurement. Our final analytic cohort contained N=407 unique women. We set each mother's first observation period as the first period in which their exposures were measured for at least 1 year, allowing for mothers to be followed for up to 5 distinct periods. We treated the outcome as a survival-type outcome such that once a woman experiences hypertension, they were considered to have it for any subsequent outcome measurement periods.  

\subsection{Exposure} \label{sec:supp:measures}
We used participant-reported addresses and California’s Pesticide Use Reporting (PUR) database to estimate average use (kg/year) of agricultural pesticides within 1 km of residences in the 2 years preceding each study visit. When index children were age 16 (2016-2018), all women were asked to recall their residential history in detail, starting at the date of conception of their index child, including move in and move out dates. Thereafter, residential history was collected prospectively at each follow-up visit.

The California Department of Pesticide Regulation’s PUR database contains application information including active ingredient, amount, date, and location,\cite{cdpr2023} which allows for the creation of geospatial measures of chronic pesticide exposure that have been found to correlate with community air samples.\cite{harnly2005correlating, madrigal2023contributions} We selected exposure periods of 2 years to estimate recent chronic exposure while minimizing potential temporal effects on the amount or type of pesticides used (e.g., seasonal or climatic changes, industry or regulatory changes). We selected a radius of 1 km because studies comparing measured pesticide concentrations in house dust or ambient air with proximity to agricultural pesticide use have observed the strongest correlations for pesticide use within 1 km.\cite{harnly2005correlating, madrigal2023contributions, harnly2009pesticides, gunier2011determinants} 

The methods used to calculate residential proximity to agricultural pesticide use have been published previously.\cite{gunier2011determinants, nuckols2007linkage, gunier2017prenatal} Briefly, we geocoded residential addresses to obtain latitude and longitude coordinates, and we used a geographic information system to create a 1-km buffer radius around each residence. Then, we calculated the kilograms of active ingredient applied within this radius by weighing the reported use within each Public Land Survey System section by the proportion of land area within the 1-km buffer. Participant-reported move-in and move-out dates were used to create continuous estimates of annual agricultural pesticide use over each 2-year exposure period for each active ingredient.

We selected seven pesticide classes based on experimental or epidemiological evidence of cardiometabolic effects: 1) organophosphate insecticides (a summary measure combining acephate, chlorpyrifos, diazinon, malathion, oxydemeton-methyl, bensulide, naled, dimethoate, methidathion, disulfoton, fenamiphos and methamidophos); 2) pyrethroid insecticides (a summary measure combining permethrin and cypermethrin); 3) carbamate insecticides (a summary measure combining carbaryl, carbofuran, cycloate, EPTC, methomyl, oxamyl and thiodicarb); 4) neonicotinoid insecticides (a summary measure combining clothianidin, dinotefuran, thiamethoxam, acetamiprid and imidacloprid); 5) manganese-containing fungicides (a summary measure combining maneb and mancozeb), 6) glyphosate herbicide; and 7) paraquat herbicide. Estimates for the individual active ingredients from each class were summed to create the final class estimates.

\section{A test of the null hypothesis of no interaction}\label{sec:supp:hypothesis} 

We can  statistically these null hypotheses of no interaction. Estimated using TMLE or SDR, the sampling distributions of $\E[Y(\bar{\mathbf{A}}^{\dd^{\bar{A6}, \bar{A7}}})]$, $\E[Y(\bar{\mathbf{A}}^{\dd^{\bar{A7}}})]$, and $\E[Y(\bar{\mathbf{A}}^{\dd^{\bar{A6}}})]$ are asymptotically normal with variance equal to the sampling variance of their respective efficient influence curves ($D_{6,7}$, $D_7$, and $D_6$). Note that this does \textit{not} mean that the outcome is normally distributed---we make no parametric assumptions. Furthermore, the nonparametric efficient influence curve of $E[Y]$ is $D_Y = Y - \E[Y]$. Then, the test statistic under the null hypothesis of no interaction is also asymptotically normal with variance equal to the sample variance of $D_{6,7} - D_7 - D_6 + D_Y$. We can therefore use $D_{6,7} - D_7 - D_6 + D_Y$ to construct a 95\% confidence interval for $\E[Y(\bar{\mathbf{A}}^{\dd^{\bar{A6},\bar{A7}}})] - \E[Y(\bar{\mathbf{A}}^{\dd^{\bar{A7}}})] -\E[Y(\bar{\mathbf{A}}^{\dd^{\bar{A6}}})] + \E[Y]$. If this confidence interval does not include zero, we can reject the null hypothesis of no interaction. The \textit{lmtp}\cite{lmtp2023} package makes this hypothesis test easy to perform as it returns objects from the \textit{ife}\cite{ife} package, which implements auto-differentiation for arithmetic operations on influence function-based estimands.

\section{Exploring low-density areas of the exposure mixture that would rely on interpolation or extrapolation}
\label{sec:bao}
\begin{figure}[H] 
  \caption{Density for baseline paraquat and glyphosate values, where threshold values separate areas where the density is greater than versus less than the threshold.}
\centering
\includegraphics[width=0.8\textwidth]{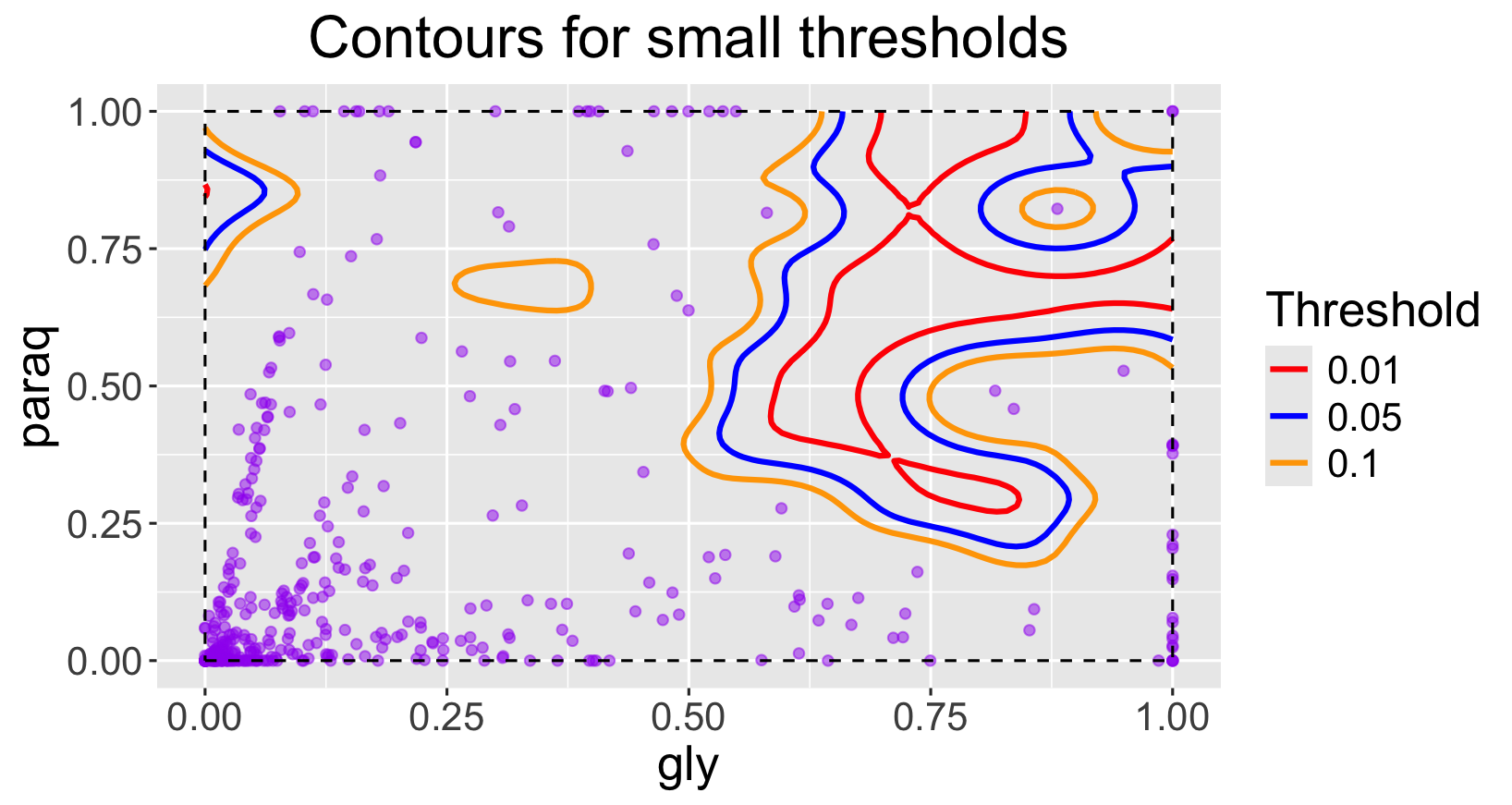}
\label{fig:supportdensity}
\end{figure}

In the main text, we assess the risk of extrapolating outside the convex hull of the marginal exposure distribution. However, even within the convex hull, there may be areas where certain values of the exposure mixture are either not observed or sparse. 
Bao and Schomaker, 2025, \cite{bao2025addressing} consider this gradation of exposure density. One can conduct an analogous examination of the environmental mixture identifying low-density regions. Figure \ref{fig:supportdensity} provides an example two-dimensional summary considering glyphosate and paraquat at baseline.

\section{Question: Over what population or subpopulation should we be estimating the relationship between the exposure mixture and outcome?}
\label{sec:localsupp}
Recall the causal and statistical estimands from Eq 4 and Eq 5, incorporating the shifts that stay within the convex hull, defined in Section 5. Each estimand is an average over the entire population---those individuals who would have their exposure values shifted because the shifted observations are supported by the data and those individuals who would not have their exposure values shifted because such a shift would be unsupported. Thus, estimating $\E(Y(\bar{\mathbf{A}}^{\dd}) -Y)$, would result in averaging over zeros for those individuals whose shifted exposure values would rely on extrapolation, which will drive the estimate towards the null and may actually not correspond to the estimand we have in mind. This issue leads us to the question: Over what population should we be estimating the relationship between the exposure mixture and outcome?

A fundamental principle of epidemiologic analysis involves restricting the analysis to those individuals who could plausibly have the exposure or treatment of interest and who are ``at-risk" of the outcome. This principle justifies refining the estimand of interest to consider those whose shifted exposure values would remain supported. 
We can thus modify Eq 4 to reflect this refined subpopulation: 
\begin{equation}
\label{eq:att}
    \E(Y(\mathbf{A}^\dd) - Y | \mathbf{A} \in \mathbb{B}),
\end{equation}
where for simplicity we only consider exposure at baseline and where $\mathbb{B}$ represents the subset of the exposure values that define this subpopulation. Note that the well-known average treatment effect on the treated (ATT) is a special case of Eq \ref{eq:att} where $\mathbb{B}$ is the set where $ A = 1$. 

We return to our running example of considering a joint reduction on both paraquat and glyphosate. Such a reduction may only be of any practical relevance if either or both of the pesticide exposures are above some minimal threshold; we set these to $\ge25$kg/yr for glyphosate and $\ge5$kg/yr for paraquat, representing the 57th and 68th percentiles of each pesticide class, respectively. This would include roughly roughly 55\% of our cohort if we require either or both thresholds to be met. Thus, we would define the causal estimand to be: $$\E(Y(\mathbf{A}^{\dd^{A6,A7}}) - Y | \mathbf{A} \in \mathbb{B}),$$
where $\mathbb{B} = (A_{6,0} \ge 25 \cup A_{7,0} \ge 5)$ 
and where 
\begin{equation}
    \label{eq:67base}
    \dd^{A6,A7}=   (A_{1,0}, A_{2,0}, ..., 0.8 \times A_{6,0}, 0.8 \times A_{7,0}).
\end{equation} We estimate a risk reduction of 2.4 percentage points (95\% CI: 0.3, 4.5) in this subgroup. 
As expected, this is appreciably greater in magnitude than our estimate of the causal estimand from Eq 4 that does not restrict to the subpopulation defined by $\mathbf{A} \in \mathbb{B}$, which is estimated to be a risk reduction of 1.4 percentage points (95\% CI: -0.6\%, 3.3\%). Extending such conditioning sets, $\mathbb{B}$, to be defined in terms of longitudinal natural values of joint exposure values has been done, but entails a less than intuitive identification assumption and interpretational challenges. The 
interested reader can consult \cite{susmann2024local}

\section{Estimator details}
\label{sec:estimatorsupp}
Both TMLE and SDR estimators are based on the so-called efficient influence curve of the statistical estimand, which is central to nonparametric estimators.\cite{kennedy2024semiparametric} 
When we say an estimator is robust to nuisance parameter misspecification, we mean that as the sample size approaches infinity the estimator is unbiased (this is referred to as being consistent). 
Although both TMLE and the SDR estimator are doubly robust, they are robust in different ways (Table \ref{tab:drprop}). TMLE is consistent if: 1) all outcome models are consistent for all timepoints, 2) all exposure and censoring (e.g., loss to follow up) models are consistent for all timepoints, or 3) for any time, $t$, all outcome models are consistent for timepoints after $t$ and all exposure and censoring models are consistent for timepoints up to and including $t$. This is in contrast to the SDR estimator, which is consistent if for all timepoints either the outcome model or exposure and censoring models are consistently estimated. 

\begin{table}[H]
    \centering
    \caption{Examples of double robustness and sequential double robustness.}
    \begin{tabular}{llccccc}
        \toprule
         & & $t=0$ & 1 & 2 & 3 & 4\\
         \midrule
         \multirow{2}{*}{Doubly robust} & Treatment & \cmark & \cmark & \xmark & \xmark & \xmark \\
         & Outcome & \xmark & \xmark & \cmark & \cmark & \cmark \\
         \midrule
         \multirow{2}{*}{Sequentially doubly robust}& Treatment & \cmark & \xmark & \xmark & \cmark & \xmark \\
         & Outcome & \cmark & \cmark & \cmark & \xmark & \cmark \\
         \bottomrule
    \end{tabular}
   \label{tab:drprop}
\end{table}

\section{Motivating data analysis}\label{sec:supp:analysis}
For the single timepoint setting, we assume observed data $O=(\mathbf{L}, \mathbf{A}, C, Y)$, where $\mathbf{L}$ is the vector of baseline covariates, $\mathbf{A}$ is the vector of exposure to 7 pesticide classes at baseline (organophosphates, pyrethroids, carbamates, neonicotinoids, maganese fungicides, glyphosate, and paraquat herbicides referred to as $A_{1}$, $A_{2}$, $A_{3}$, $A_{4}$, $A_{5}$, $A_{6}$, $A_{7}$, respectively), $C$ indicates censoring, meaning that the individual did not come to the follow-up visit when the outcome was measured, and $Y$ is the outcome of hypertension by the fifth follow-up timepoint, observed 10 years after baseline. We are interested in the causal estimand: $\theta^* = \E(Y(\mathbf{A}^{\dd^*}, C=1) - Y(C= 1) )$, for each $\dd^* \in \{\dd^{\text{all}}, \dd^{\text{first}5}, \dd^{A6,A7}, \dd^{A7}\}$, where $\dd^{\text{all}}$ is defined in Eq 1, $\dd^{\text{first}5}(\mathbf{A}) = 
 (0.8 \times A_{1}, ..., 0.8 \times A_{5}, A_{6}, A_{7})$, $\dd^{A6,A7}$ is defined in Eq \ref{eq:67base}, and $\dd^{A7}(\mathbf{A})=(A_{1}, ..., 0.8 \times A_{7})$. 
In words,  $\theta^{\text{first}5}$ is the expected difference in risk of developing chronic hypertension by the fifth follow-up visit had all mothers in the CHAMACOS cohort had their exposure levels of organophosphates, pyrethroids, carbamates, neonicotinoids, and maganese fungicides reduced by 20\% at baseline (leaving their levels of glyphosate and paraquat as observed), as compared to their observed hypertension risk by the fifth follow-up visit. $\theta^{all}$, $\theta^{A6,A7}$ $\theta^{A7}$ are defined similarly.



As for the longitudinal analysis, we used a targeted minimum loss-based estimator with 20 cross-fit folds to estimate the effects.\cite{diaz2023nonparametric} 
We estimated outcome nuisance parameters using a stacked ensemble\cite{van2007super} of: an intercept-only model, a main-effects generalized linear model, LASSO,\cite{tibshirani1996regression} multivariate adaptive regression splines\cite{friedman1991multivariate}, random forests\cite{breiman2001random}, and several implementations of light gradient boosting\cite{ke2017lightgbm}. 
Analyses were conducted using R (version 4.5.1)\cite{Rcitation} with the \textit{lmtp} \cite{lmtp2023}, \textit{mlr3superlearner}\cite{mlr3superlearner}, and \textit{torch}\cite{Rtorch} packages. All code for the analysis is available at \url{https://github.com/CI-NYC/CHAMACOS/blob/main/scripts/3a_analysis_single_timepoint.R}

\begin{figure}[H]
 \caption{Effect estimates of 20\% reductions in pesticide exposure at baseline on risk of hypertension.}
 \centering
\includegraphics[width=1\textwidth]{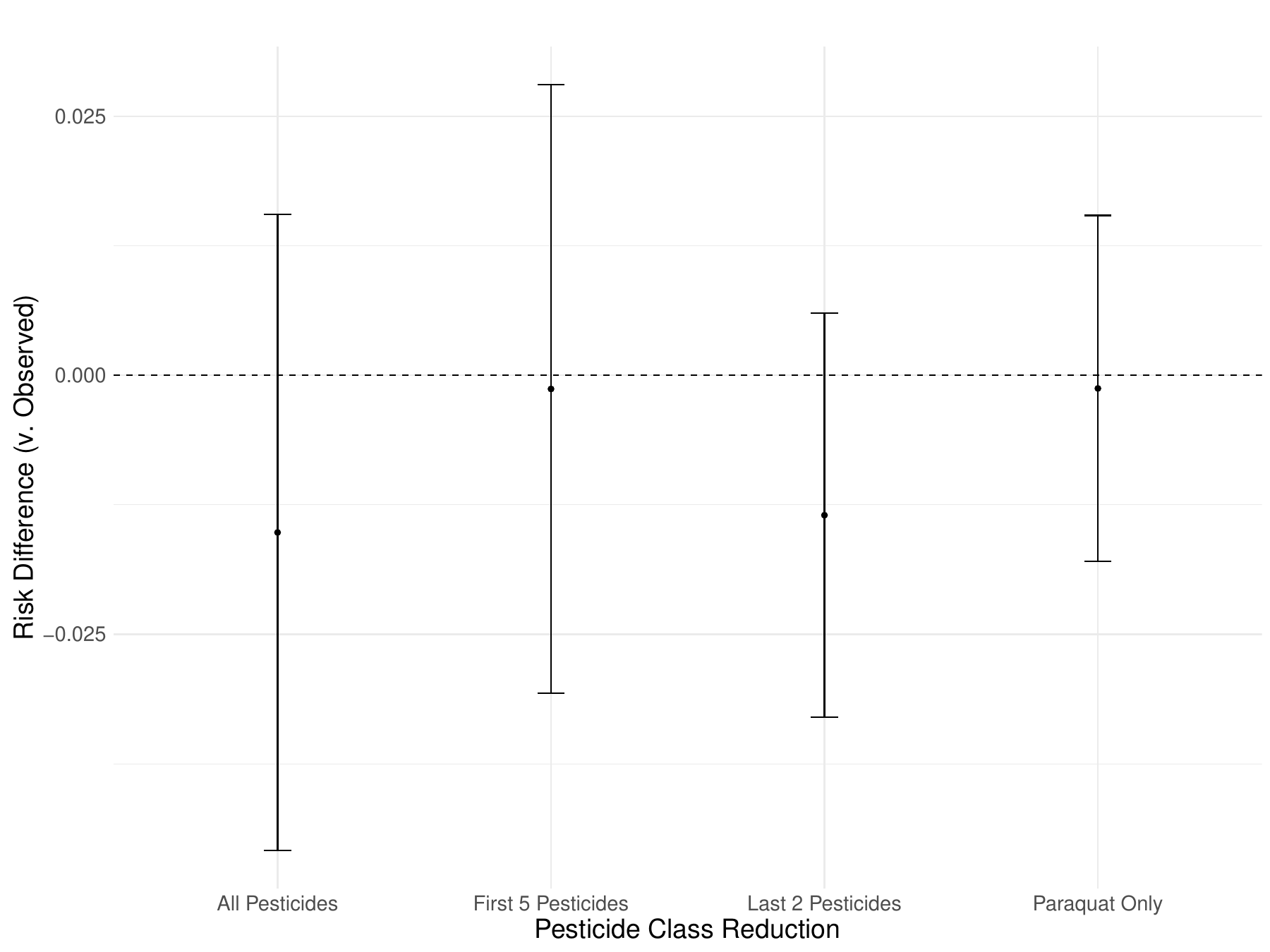}
\label{fig:single}
\end{figure}

We show estimates of each $\hat \theta^*$ in Figure \ref{fig:single}. We estimate small, nonsignificant effects of reducing pesticides at a single timepoint on hypertension risk at the final followup point. In the single timepoint setting, we estimate that reducing all pesticide classes by 20\% at baseline would be expected to reduce risk of hypertension by the last follow-up visit by 1.5 percentage points (95\% CI: -1.6, 4.6) (see Figure \ref{fig:single}). Reducing just the first 5 pesticide classes (risk reduction of 0.1 percentage points, 95\% CI: -2.8\%, 3.1\%), the last 2 pesticide classes (risk reduction of 1.4 percentage points, 95\% CI: -0.6\%, 3.3\%), or paraquat only (risk reduction of 0.1 percentage points, 95\% CI: -1.5\%, 1.8\%) at baseline yielded smaller reductions in risk of hypertension to varying extents.

\bibliography{refs}
\end{document}